\documentclass[aps,prb,twocolumn,amsmath,superscriptaddress]{revtex4-2}

\newcommand{\bea}{\begin{eqnarray}}
\newcommand{\eea}{\end{eqnarray}}
\newcommand{\beq}{\begin{equation}}
\newcommand{\eeq}{\end{equation}}

\newcommand{\dd}{d^{\dagger}}
\newcommand{\dnd}{d^{\phantom{\dagger}}}

\usepackage[urlcolor=blue,colorlinks=true,citecolor=blue,linkcolor=blue,pdfstartview={FitH},bookmarks=false]{hyperref}
\usepackage{graphicx}
\usepackage{longtable}
\usepackage{epsfig}
\usepackage{dcolumn}
\usepackage{bm}
\usepackage{amssymb}
\usepackage{multirow}
\usepackage{times,xcolor}
\usepackage{hyperref}
\usepackage{braket}
\usepackage{comment}

\begin{document}

\title{Screening in a two-band model for superconducting infinite-layer nickelate}

\author{Tharathep Plienbumrung}
\author{Maria Daghofer}
\affiliation{\mbox{Institute for Functional Matter and Quantum Technologies,
University of Stuttgart, Pfaffenwaldring 57, D-70550 Stuttgart, Germany}}
\affiliation{\mbox{Center for Integrated Quantum Science and Technology,
University of Stuttgart, Pfaffenwaldring 57, D-70550 Stuttgart, Germany}}

\author{Michael Schmid}
\affiliation{\mbox{Waseda Research Institute for Science and Engineering,
Waseda University, Okubo, Shinjuku, Tokyo, 169-8555, Japan}}

\author{Andrzej M. Ole\'s$\,$}
\email{Corresponding author: a.m.oles@fkf.mpi.de}
\affiliation{\mbox{Max Planck Institute for Solid State Research,
             Heisenbergstrasse 1, D-70569 Stuttgart, Germany} }
\affiliation{\mbox{Institute of Theoretical Physics, Jagiellonian University,
             Profesora Stanis\l{}awa \L{}ojasiewicza 11, PL-30348 Krak\'ow, Poland}}

\begin{abstract}
Starting from an effective two-dimensional two-band model for infinite
layered nickelates, consisting of bands obtained from $d$ and $s$--like
orbitals, we investigate to which extend it can be mapped onto a
single-band Hubbard model. We identify screening of the more itinerant
$s$-like band as an important driver. In absence of screening one
strongly-correlated band gives an antiferromagnetic ground state. For weak
screening, the strong correlations push electrons out of the $s$-band so that
the undoped nickelate remains a Mott insulator with half filled $d$ orbitals.
This regime markedly differs from the observations in high-$T_c$ cuprates and
pairing with $s$-wave symmetry would rather be expected in the
superconducting state. In contrast, for strong screening, the $s$ and
$d_{x^2-y^2}$ bands are both partly filled and couple only weakly, so
that one approximately finds a self-doped $d$ band as well as tendencies
towards $d$-wave pairing. Particularly in the regime of
strong screening mapping to a one-band model gives interesting spectral
weight transfers when a second $s$ band is also partly filled.
We thus find that both one-band physics and a Kondo-lattice--like regime
emerge from the same two-orbital model, depending on the strength of
electronic correlations and the size of the $s$-band pocket.
\end{abstract}

\date{\today}

\maketitle


\section{Introduction}

The recent discovery of superconductivity in infinite-layer NdNiO$_2$
thin films with Sr doping \cite{Li19} has rekindled interest in
Ni-based superconductivity. The interest arises mostly because they
can be assumed to be strongly correlated and thus in some respect
similar to cuprate superconductors.
The two families of compounds also have similar lattices, with either
CuO$_2$ or NiO$_2$ planes that give them a predominantly two-dimensional
(2D) character. Analyzing the NiO$_2$ layer in analogy to a CuO$_2$
layer, one expects Ni$^{1+}$ with a $d^9$ electronic configuration,
and in both cases expects antiferromagnetic (AF) superexchange
interactions~\cite{Jia20,Tha21,Hep20}. Indeed, magnetic excitations
in undoped nickelates reveal such AF correlations~\cite{Lu21,Lin21}.

However, the parent compound $R$NiO$_2$ ($R=$La, Nd)
does not show any signs of magnetic ordering \cite{Hay99,Hay03} at low
temperatures down to 1.5 K. Moreover, both the insulating NiO$_2$
layer~\cite{Jia20,Tha21,Hu19,Zha20} and band-structure calculations
\cite{Jia19,Si20,Wu20,Kle21,Zha21,Bee21,Hig21,Bot22} suggest that other
orbitals or bands might be relevant.

\begin{figure}[t!]
	\includegraphics[width=\linewidth]{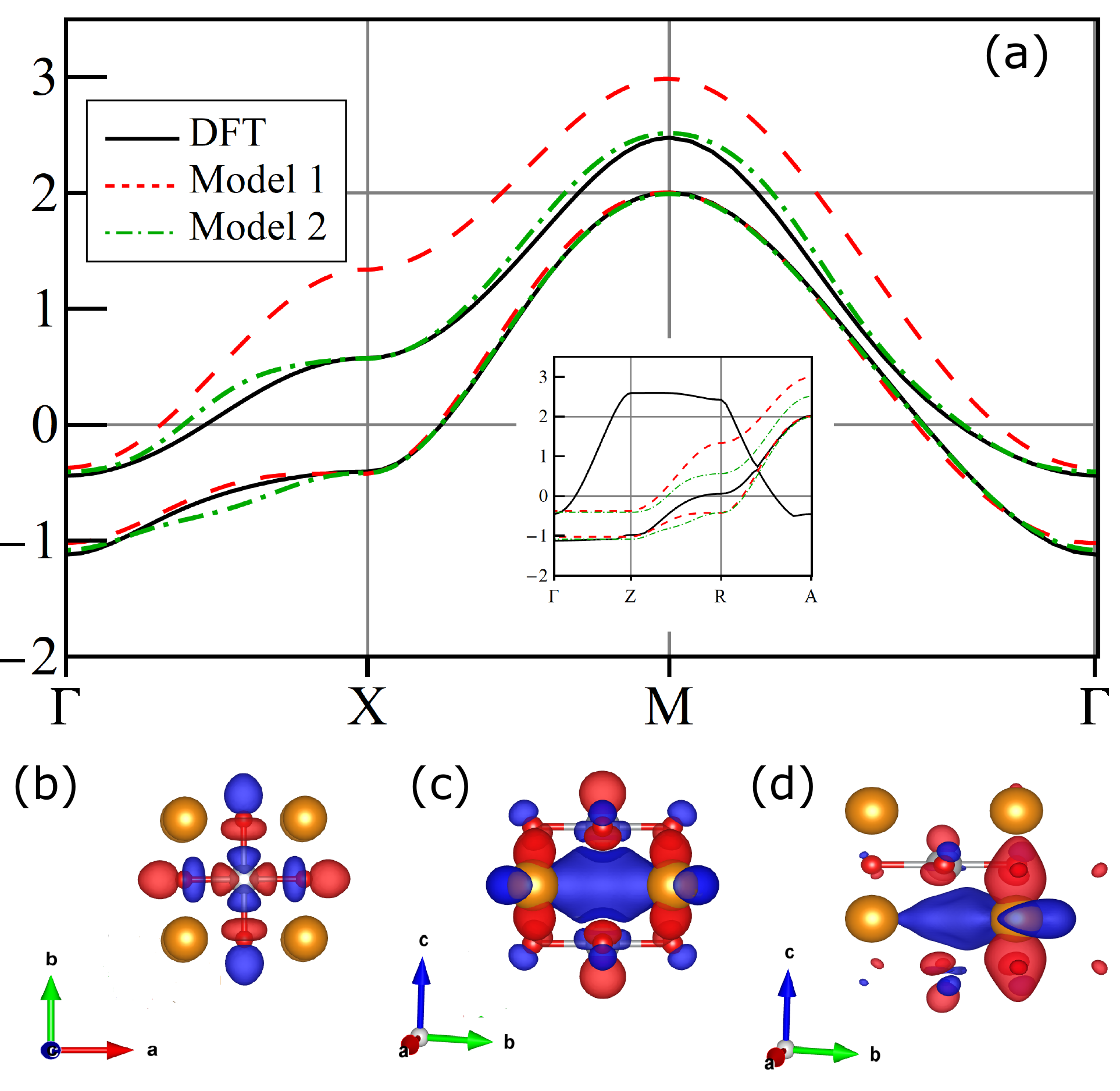}
\caption{\textcolor{black}{Non-interacting band structure of NdNiO$_2$:
(a) Density-functional theory (DFT) bands crossing the Fermi surface
    (black) and two 2D tight-binding models obtained by projecting a Wannier
    fit onto the plane along  2D path (red dashed and green dotted).
    The Fermi energy $E=\mu$ corresponds to the DFT electronic structure.
    Inset shows the DFT band structure along three-dimensional path
    $\Gamma-Z-R-A$. Wannier orbitals are also shown: (b) The
    '$d_{x^2-y^2}$-like' orbital making up the lower band of model 1;
    the corresponding orbital of model 2 looks the same. The
    '$s$-like' orbitals of models 1 and 2 are given in (c) and (d).}}
	\label{fig:bands}
\end{figure}

Several approaches suggest that a two-band model should be a realistic
starting model for doped infinite-layer nickelates
\cite{Adh20,Nom19,Gu20,LecX,Xie21}. It faithfully represents the
strongly-correlated $x^2-y^2$ orbital at Ni ions and $s$ orbital which
collects all remaining contributions from other orbitals. Indeed, two
bands cross the Fermi energy as shown in Fig. \ref{fig:bands}.
Accordingly, a large variety of potential pairing symmetries has been
presented \cite{Hu19}, with exotic $s+id$ \cite{Wan20p} states proposed
in addition to $s$ and $d$-wave pairings \cite{Wu20,Zha20}. However,
the majority of models, originating from one~\cite{Kit20} to
three~\cite{Wu20} bands, identify $d$-wave pairing \cite{Nom21} as in
cuprates. Experimentally, $d$-wave symmetry was reported, as was
$s$-wave~\cite{Gu20a}.

Microscopically, two main differences between Cu- and Ni-based
superconductors are: (i) larger Ni-O charge-transfer energy
compared to Cu-O, and (ii) the presence of highly dispersive rare-earth
bands in the layered nickelates~\cite{Bot20,Hep20}. In cuprates, doped
holes go mostly into oxygen sites, where they form Zhang-Rice singlets
with the half-filled Cu($d_{x^2-y^2}$) states~\cite{Zha88}. Due to the
larger charge-transfer energy raising the energy of the oxygen orbitals,
Ni--O hybridization is less important and doped holes likely reside on
Ni sites in doped Nd$_{1-x}$Sr$_x$NiO$_2$
\cite{Jia20,Sak20,ZYZ20,Tha21,Ros21}. While oxygen states can thus be
assumed to play less of a role~\cite{Kit20},
the question of the orbital character of doped holes remains and is
affected by the rare-earth band. This band hybridizes with Ni apical
states, thus obtaining some Ni($d_{3z^2-r^2}$) and Ni($d_{xy}$)
character, and forms a hybrid 'axial' $s$ orbital~\cite{Adh20}, see
Fig.~\ref{fig:bands}. Depending on where doped holes go, one can expect
either that almost only the $d_{x^2-y^2}$ states are relevant
\cite{Kit20}, or that half-filled more localized $d_{x^2-y^2}$ states
together with itinerant $s$-like carriers form a Kondo-lattice--like
two-band system \cite{Hep20,Wan20p,ZYZ20,Bee21}, or that both scenarios
have their point~\cite{Lec21}.

The purpose of this paper is to investigate how similar cuprate and
nickelate superconductivity are, i.e., whether and where multi-band
effects come into play. We find that screening is a crucial variable
and that the two bands (Fig. \ref{fig:bands}) can mix considerably at
intermediate screening, where pairing with $s$-wave symmetry would be
expected. Both for very weak and for strong screening, however, the
bands mostly decouple and the effective physics becomes similar to a
single Hubbard band~\cite{Kit20}. We then find Mott insulator (doped
band with potential $d$-wave pairing) for strong (weak) correlations.

The remaining of this paper is organized as follows. The two-band model
arises from the electronic structure calculations as described in Sec.
\ref{sec:two}. Electronic interactions are given by two Kanamori
parameters $\{U_{\alpha},J_H\}$ and we discuss their screening in Sec.
\ref{sec:int}. In Sec. \ref{sec:num} the results of exact
diagonalization are presented for the density distribution and spin
correlations (Sec. \ref{sec:dos}) and for the spectral density (Sec.
\ref{sec:one}). We search for superconducting (SC) phases both in
hole doped and electron-doped systems in Sec. \ref{sec:sup}. Next we
present phase diagrams of infinite-layer
nickelates in $(U_d,\alpha)$ planes in Sec. \ref{sec:phd}. Two SC
compounds, NdNiO$_2$ and LaNiO$_2$ are compared in
Sec. \ref{sec:la}. The paper is concluded in Sec. \ref{sec:summa}.

\section{Two-band model and Methods}
\label{sec:mod}
\subsection{Kinetic energy}
\label{sec:two}

We start from the kinetic energy in the electronic structure. The DFT
band structure, see Fig. \ref{fig:bands}, is calculated with
{\sc Quantum} {\sc Espresso} code \cite{Gia09,Gia17,Gia20} using a
plain-wave pseudopotential method; similar calculations were performed
previously \cite{Adh20,Bot20,Sak20}. In Fig.~\ref{fig:bands}(a), one
finds that two bands cross the Fermi level. Wannier-orbital models were
obtained to reproduce the two bands crossing the Fermi level $\dots$.
In both cases, the Wannier orbital corresponding to the lower band
contains substantial Ni $d_{x^2-y^2}$ contributions, with some weight
on the surrounding O atoms, see Fig.~\ref{fig:bands}(b). The upper band
is formed by a rather extended illustrated, where
Nd($5d$) orbitals hybridize with Ni($s$) as well as Ni($d_{xy}$) and
Ni($d_{3z^2-r^2}$) states. \textcolor{black}{Both are compromises to
some extend: one of the fits (model 2) has a smaller imaginary part, but
the orbitals of the other (model 1) more closely respect the expected
symmetries, see Fig.~\ref{fig:bands}(c) and \ref{fig:bands}(d).
Hoppings are, however, very similar.}

The feature of the band structure that most distinguishes nickelates
from cuprates, are the electron pockets formed by the upper $s$-like
band around the $\Gamma$ and $A$ points in the Brillouin zone.
In the DFT band structure, they contain $\approx7\,\%$ of the occupied
states~\cite{Bot20}, which in turn implies that there are hole carriers
\cite{ZYZ20} in the $x^2-y^2$ band even without Sr-doping. While
$7\,\%$ self doping may not seem much, it is in line with the $5\,\%$
of Sr doping needed to destroy antiferromagnetism in a cuprate
superconductor La$_2$CuO$_{4-y}$ \cite{Bud88}. The $\Gamma$-pocket
lying about $-0.4$ eV below the Fermi level appears thus to be an
important feature when constructing an effective model.

The Wannier90 interface~\cite{Piz20} gives the parametrization
\begin{align}
\label{eq1}
H_{\rm kin}&=\sum_{i\alpha\sigma}\epsilon_{\alpha}
\dd_{i\alpha\sigma}\dnd_{i\alpha\sigma}
+\sum_{ij\alpha\beta\sigma}{t_{ij}^{\alpha\beta}
d^{\dagger}_{i\alpha\sigma}d^{}_{j\beta\sigma}},
\end{align}
of these two bands, where $\dnd_{i\alpha\sigma}$ ($\dd_{i\alpha\sigma}$)
is an electron annihilation (creation) operator at site $i$ for orbital
$\alpha$ and spin $\sigma$. $\alpha=d$ denotes the Ni-$x^2-y^2$
dominated state of Fig.~\ref{fig:bands}(b) and $\alpha=s$ stands for the
extended $s$-like state of Fig.~\ref{fig:bands}(c). Hopping parameters
$t^{\alpha\beta}_{ij}$ and on-site energies $\epsilon_{\alpha}$ are
given in the Supplemental Material (SM) \cite{suppl}
\textcolor{black}{for two slightly different Wannier projections}.

With exact diagonalization and the Lanczos algorithm \cite{Koch},
we can address an eight-site cluster,
e.g. the $2\times2\times2$ cluster that was used to investigate
magnetic order~\cite{Tha21a}. We are here mostly interested in
electronic correlations, which predominantly affect the $d$ band
(see discussion below). As this band has nearly no dispersion along
the $z$-direction, see inset of Fig.~\ref{fig:bands}(a), we use instead
a $\sqrt{8}\times\sqrt{8}$ cluster in the $(x,y)$ plane.
We thus need a 2D projection of the band structure, which we obtained
by a fit that can in turn be motivated by twisted boundary conditions
(TBC)~\cite{Shi97,Poi91,Yin09}. The resulting 2D bands are shown in Fig.
\ref{fig:bands}(a), hoppings parameters and details of the procedure
are given in the SM~\cite{suppl}. \textcolor{black}{This was done for both
Wannier projections, leading to two similar effective two-dimensional
two-band models, whose main differences concern the upper band. As can
be seen in the SM~\cite{suppl}, results of both models are very similar,
so that we focus the main part of the paper on the model 1.}

\subsection{Interactions and screening effect}
\label{sec:int}

Electronic interactions are taken to be onsite, i.e.,
\begin{align}
\label{eq2}
\!H_{\rm int}&=
\sum_{i\alpha}U_{\alpha}n_{i\alpha\uparrow}n_{i\alpha\downarrow}
 + \left(U'-\frac{J_{H}}{2}\right)\sum_{i}{n_{id}n_{is}} \\
& - 2J_{H}\sum_{i}\vec{S}_{id}\!\cdot\!\vec{S}_{is}
+\!J_H\sum_i\!\left(\dd_{id\uparrow}\dd_{id\downarrow}
\dnd_{is\downarrow}\dnd_{is\uparrow}+{\rm H.c.}\right)\!.   \nonumber
\end{align}
$n_{i\alpha\sigma}^{}$ is the electron number operator at site $i$,
orbital $\alpha$ and spin $\sigma$ and $\{\vec{S}_{i\alpha}\}$ is
the corresponding spin operator. Intraorbital Coulomb repulsion
$U_{\alpha=s/d}$ depends on the band $\alpha$, Hund's exchange $J_H$
and interorbital repulsion $U'$ couple the bands.

Upper limits for the 'bare' $U_d$ and $J_0$ are given by their atomic
values $U_d\approx 8$ eV and $J_0\approx 1.2$ eV, which would be
applicable to in a NiO$_2$ model for an insulating layer
\cite{Jia20,Tha21}. However, even though Ni--O hybridization is expected
to be weaker than in cuprates (due to larger Ni--O crystal-field
splitting), it is still present, see the orbital wave function in
Fig.~\ref{fig:bands}(b), and expected to substantially reduce effective
values~\cite{Zha88,Tha21}.
Note that $U_s$ cannot be related to atomic values for Ni, as the
$s$-orbital is not even centered on an Ni site~\cite{Adh20}. Coulomb
interactions are thus expected to be weaker in the $s$-like band because
the wave function is far more extended and largely of Nd $5d$ character,
see Fig.~\ref{fig:bands}(c). While the Coulomb repulsion $U_d$ is almost
unscreened in the correlated $x^2-y^2$ orbitals, considerable screening
occurs for $s$ orbitals. One thus expects $U_d> U_s$ and we introduce
here parameter $\alpha \in [0,1]$ so that
\begin{align}\label{eq:U_J_alpha}
	U_s = \alpha U_d,\quad	J_H = \alpha J_0,\quad 	U' = U_s - 2J_H.
\end{align}

\textcolor{black}{Approaches like the constrained random-phase
  approximation might provide estimates for the screened interaction
  parameters, however, this is not straightforward. Even though, as
  shown above, the two Wannier-projection schemes used above lead to
  very similar band structures, and thus hopping integrals, the orbital
  wave function of the upper band differs substantially. This would in
  turn affect effective interactions, so that we opt here for a model
  approach, where we investigate which physics can be expected for
  various screening scenarios. The parametrization of
  Eq.~(\ref{eq:U_J_alpha}) is used here as the simplest approach
  capturing the essential features of the electronic structure, i.e.,
  the interplay of a more and a less correlated band.}

The full Hamiltonian $H=H_{\rm kin}+H_{\rm int}$ thus describes a
correlated 2D band, which is strongly reminiscent of cuprates, but that
moreover interacts with a more itinerant rare-earth band. By accepting
some electrons, the itinerant $s$ band not only dope the correlated $d$
band, but also contains itinerant carriers~\cite{ZYZ20,Adh20,Sak20}.
We are next going to investigate the impact of these carriers and their
remaining correlations.

\section{Numerical results}
\label{sec:num}

\subsection{Density of states and spin correlations}
\label{sec:dos}

\begin{figure}[t!]
	\includegraphics[width=\columnwidth]{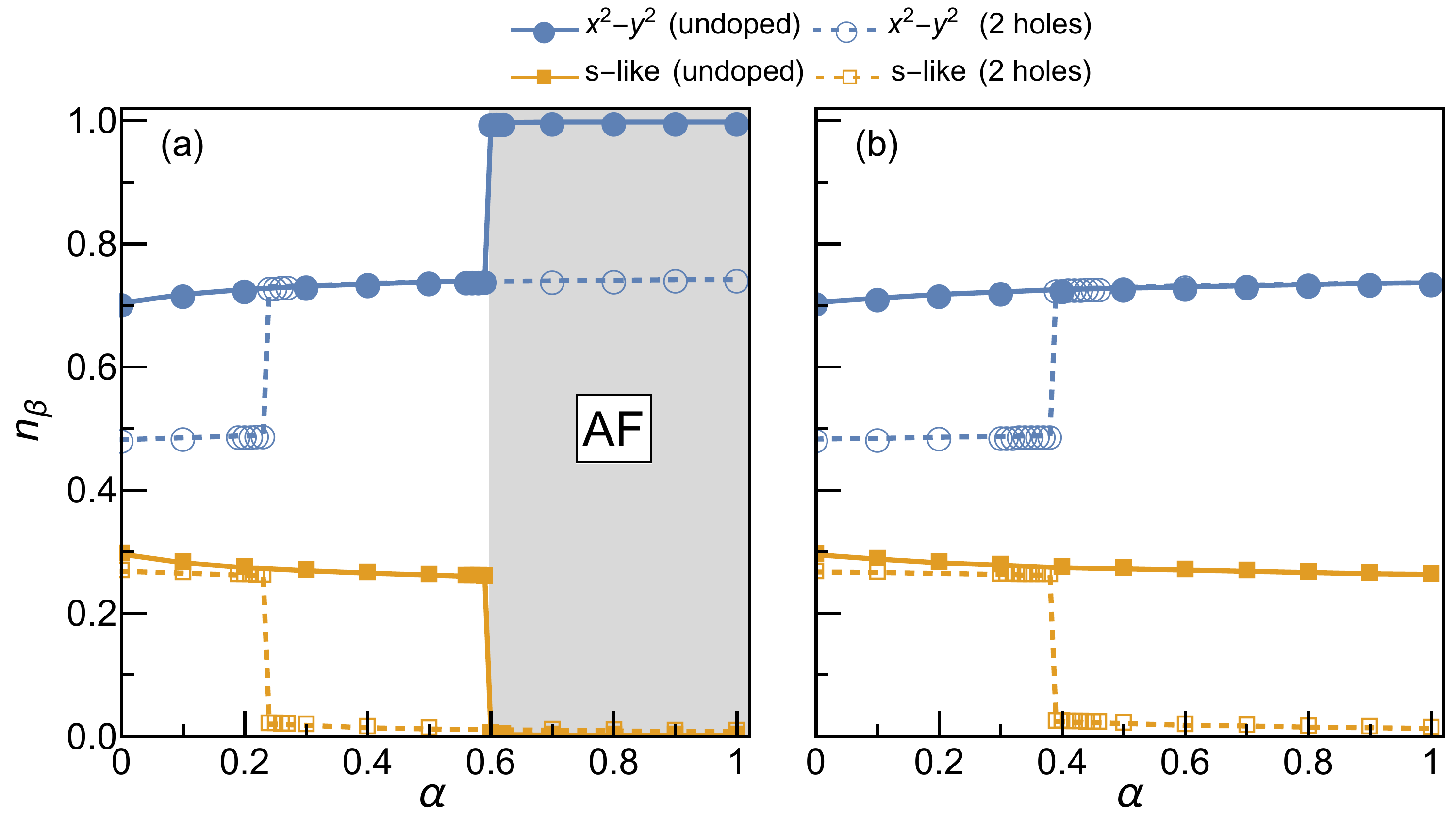}
        \caption{
          Orbital electron density \eqref{orb} as a function of screening
          parameter $\alpha$ with
          $J_0/U_d=0.15$ for
          (a) $U_d=8.0$ and $J_0=1.2$;
          (b)~$U_d=4.0$ and $J_0=0.6$.
          Solid (dashed) line for eight (six) electrons, i.e., undoped
          and doped with two holes in Ni$_8$O$_{16}$ unit.
	\label{fig:density}}
\end{figure}

\begin{figure}[b!]
	\includegraphics[width=\columnwidth]{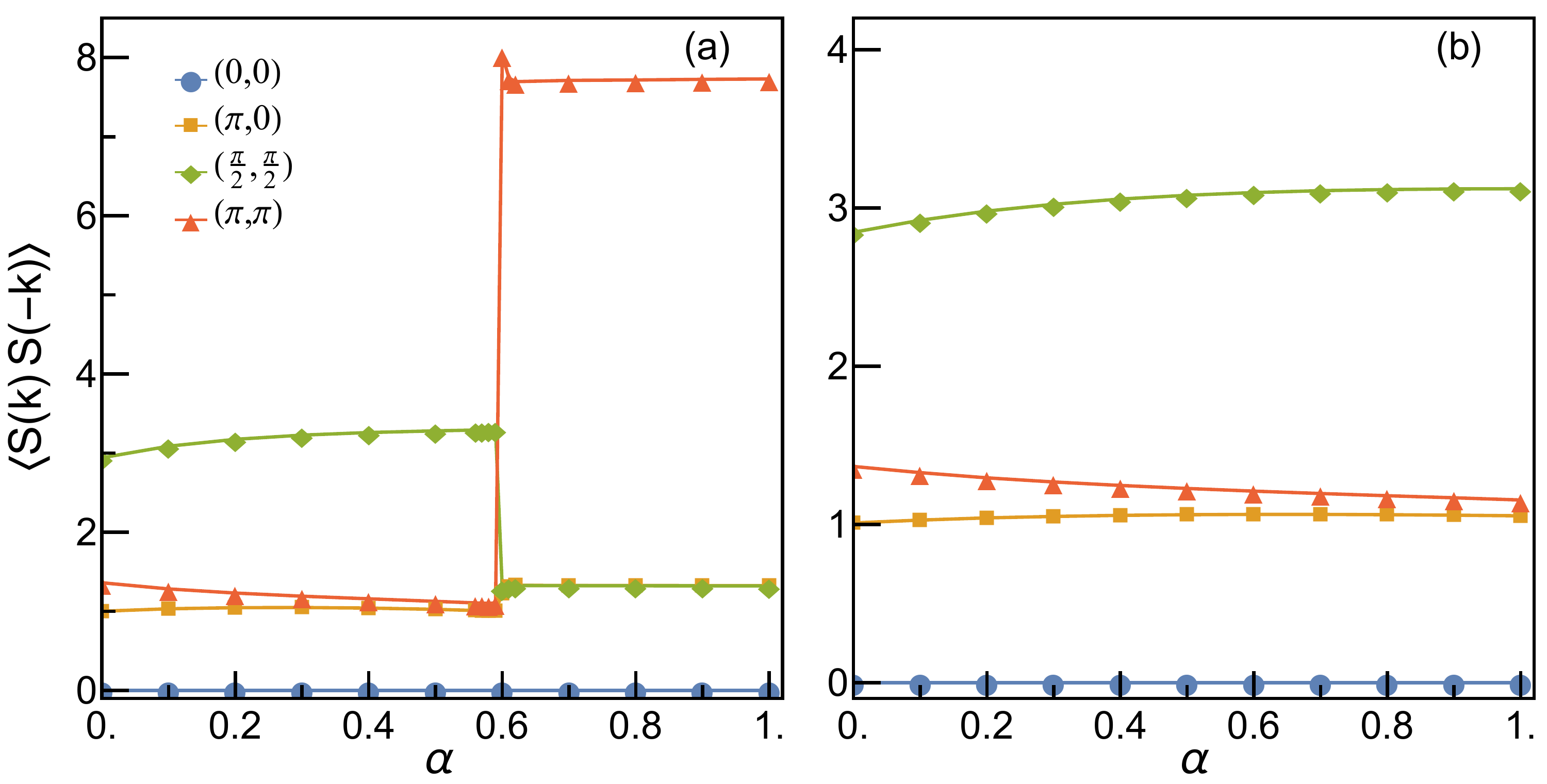}
	\caption{
		Spin structure factor of undoped system as a function of screening
		parameter $\alpha$ with
		$J_0/U_d=0.15$ for
		(a) $U_d=8.0$ and $J_0=1.2$;
		(b)~$U_d=4.0$ and $J_0=0.6$.}
	\label{fig:ssk}
\end{figure}

First we look at orbital densities,
\begin{equation}
\label{orb}
n_{\alpha}=\frac{1}{N_s}\sum_{i\sigma}
\left\langle\dd_{i\alpha\sigma}\dnd_{i\alpha\sigma}\right\rangle,
\end{equation}
where $N_s=8$ is the number of lattice sites, $\alpha=s,d$, and the
average is obtained for the ground state with 8 (6) electrons.
In the two-band model the undoped compound corresponds to
quarter-filling, i.e., eight electrons.
For very strong interactions,
band dispersion is suppressed and onsite energies dominate, so that
the $x^2-y^2$ orbital becomes half-filled. This is indeed observed for
strong $U_d=8$ eV and $J_0=1.2$ eV, see Fig. \ref{fig:density}(a).
The spin-structure factor is here strongly peaked at $(\pi,\pi)$, see
Fig. \ref{fig:ssk}(a), so that we recover the familiar picture of a
half-filled AF Mott insulator~\cite{Nom20}. However, we are in the
metallic regime and finite electron density is found in the $s$-orbital
for the uncorrelated band structure of Fig.~\ref{fig:bands}(a).

Figure \ref{fig:density} shows the orbital-resolved density versus
screening parameter $\alpha$. Here we interpolate between the strongly
correlated and uncorrelated regimes, while keeping a physically
plausible hierarchy of interactions: $U_s=\alpha U_d<U_d\le 8$ eV.
(For the moment, we keep the ratio $J_0/U_d=0.15$ constant.)
In Fig.~\ref{fig:density}(a), $U_d$ is at
its upper limit $8$ eV, and for weak to moderate screening, we find a
half-filled $x^2-y^2$ orbital with AF order. However, as soon as $U_s$
is screened by about $40\%$, self-doping occurs and some electrons enter
the $s$-band. The same happens for---presumably more realistic---$U_d=4$
eV, even without additional screening. The presence of the second band
thus causes the loss of long-range magnetic order even for parameter
regimes where the $x^2-y^2$ orbital by itself would lead to an insulator.

The next question to ask is where holes doped into the quarter-filled
system go. Three regimes emerge, see Fig.~\ref{fig:density}.
First, in the weakly screened Mott insulator and for large $U_d$, the
state is AF and holes naturally enter only the $x^2-y^2$ orbital. In
contrast, the second regime is found at intermediate screening
[$\alpha\simeq 0.5$, see Fig.~\ref{fig:density}(a)], or for interactions
that are reduced from the outset, see Fig.~\ref{fig:density}(b).
Finally, in the third regime of strong screening ($\alpha<0.5$), hole
doping happens again into the $x^2-y^2$ orbital, with the $s$ electrons
remaining unaffected.

Figure~\ref{fig:ssk} shows the spin-structure factor of undoped
nickelate at four momenta $k$ accessible to the
$\sqrt{8}\times\sqrt{8}$-site cluster. In the unscreened limit, see
Fig.~\ref{fig:ssk}(a), the AF wave vector $(\pi,\pi)$ dominates for
$\alpha>0.6$, highlighting AF Mott insulator in the strongly interacting
limit. Upon decreasing $\alpha<0.6$, the strongest signal is found at
$(\frac{\pi}{2},\frac{\pi}{2})$, but its dominance is far less
pronounced. Similarly, in Fig.~\ref{fig:ssk}(b), the strongest signal at
$(\pi,\pi)$ is suppressed once interactions in the $x^2-y^2$ orbital are
weaker.

\subsection{One-particle spectral density}
\label{sec:one}

\begin{figure}[b!]
	\includegraphics[width=\columnwidth]{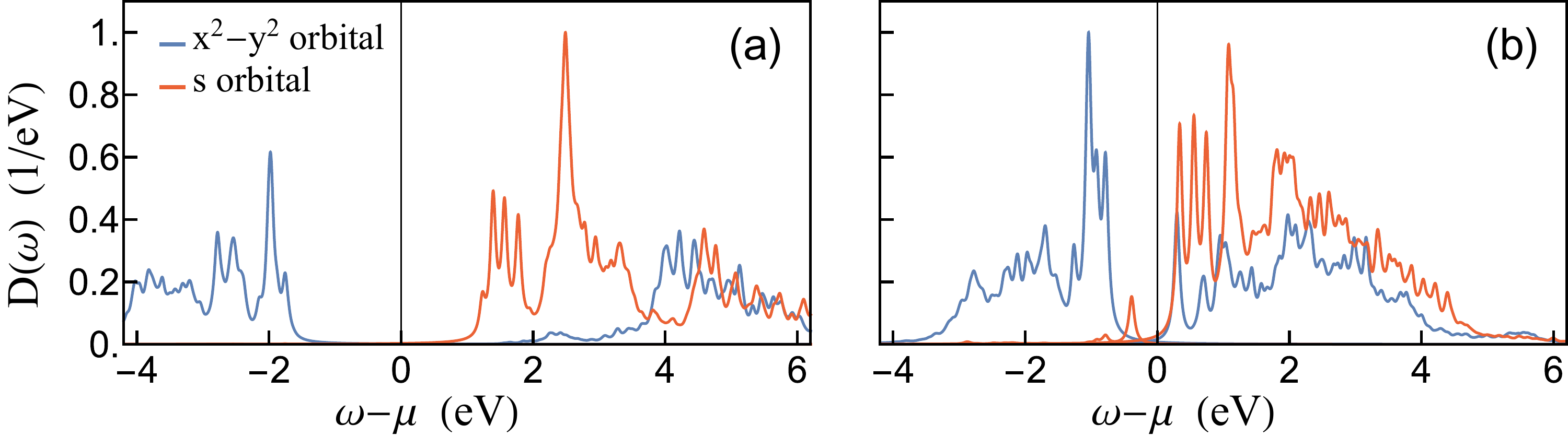}
	\caption{
Density of states $D(\omega)$ of undoped nickelate with unscreened
interactions ($\alpha=1$). Fermi energy is set to zero;
the densities are normalized to one (per spin).
Intraorbital Coulomb interaction in Eq. \eqref{eq2} is selected at:
(a) $U_d=8$ eV and
(b) $U_d=4$ eV.}
	\label{fig:mi}
\end{figure}

To understand the occurrence of possible SC phase in infinite-layer
nickelates we consider first the one-particle spectra in the normal
phase. When interactions are unscreened ($\alpha=1$), the undoped system
is a Mott insulator for $U_d\ge 4$ eV, see Fig. \ref{fig:mi}. For
$U_d=8$ eV one finds that the correlated $d$ band is half-filled and a
broad gap $\sim 3.5$ eV separates occupied from empty states, with the
Fermi energy within the gap, see Fig. \ref{fig:mi}(a). The gap in the
correlated band consisting of $x^2-y^2$ orbitals is close to 6 eV and
unoccupied states above the Fermi level are the $s$ band states. This
electronic structure corresponds to an AF Mott insulator.

When $U_d=4$ eV, the gap in the correlated band decreases to
$\sim 1.0$ eV but the tail of the $s$ band falls below the Fermi energy
which still separates the occupied and unoccupied states, see Fig.
\ref{fig:mi}(b). However, for this value of $U_d$, we cannot exclude a
metallic phase, with a small fraction of electrons occupying the $s$
states in the thermodynamic limit.

\begin{figure}[t!]
	\includegraphics[width=\columnwidth]{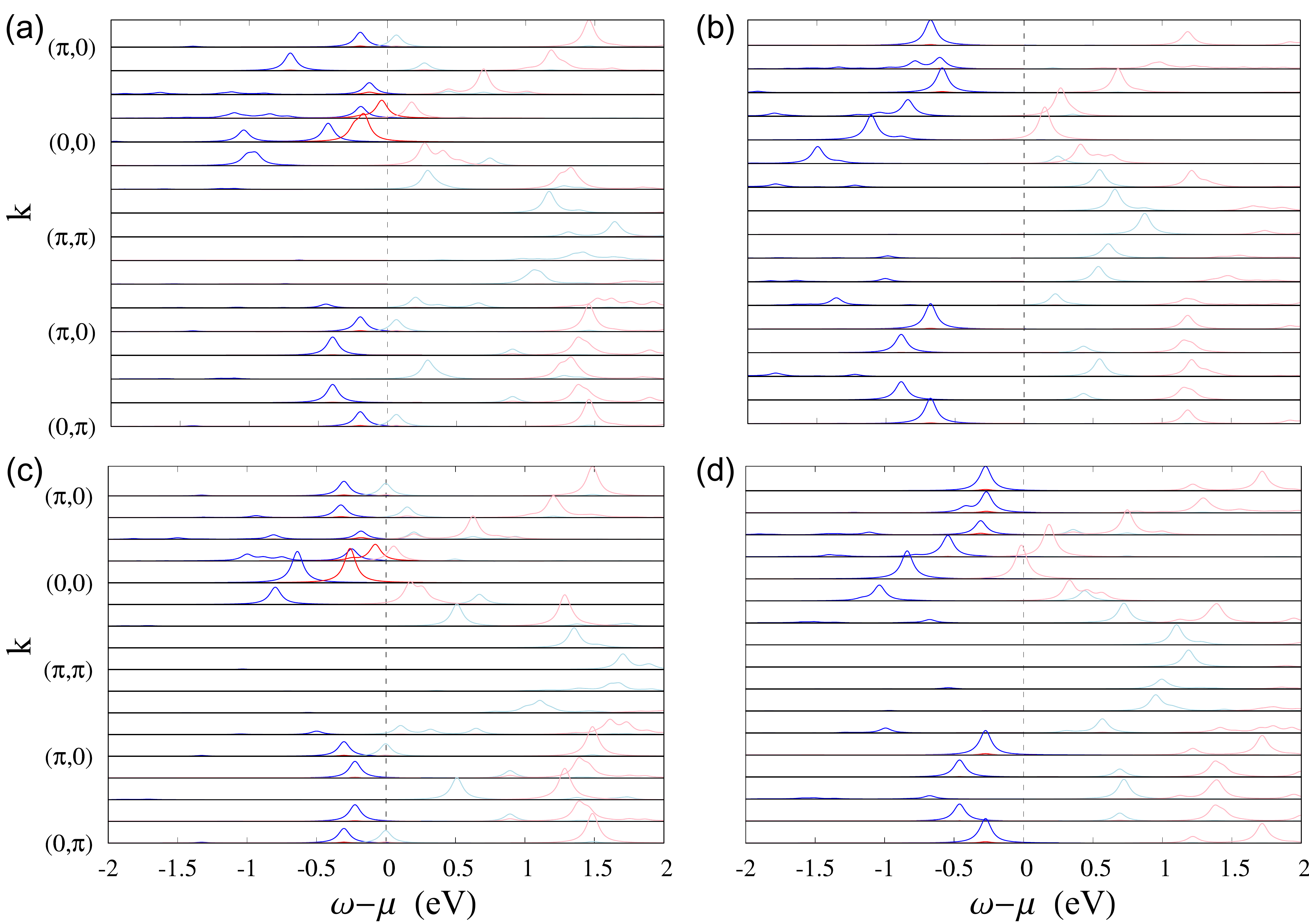}
        \caption{Single-particle spectra for six-electron system
          obtained with TBC and $J_0/U_d=0.15$:
   top---$U_d=8$ eV and (a) $\alpha=0.2$, \mbox{(b) $\alpha=0.5$;}
bottom---\mbox{$U_d=4$} eV and (c) $\alpha=0.2$, (d) $\alpha=0.5$.
Blue (red) spectra below $\mu$ correspond to occupied $x^2-y^2$ and $s$
states, respectively. Light spectra above $\mu$ are for empty states in
both bands.
	}
	\label{fig:spectra}
\end{figure}

The different behavior in the three regimes mentioned in Sec.
\ref{sec:dos} is also reflected in the single-particle spectra shown in
Fig.~\ref{fig:spectra}. Filling corresponds to doping with two holes
and TBC is used to resolve more momenta \cite{Shi97,Poi91,Yin09}. Both
for very strong $U_d=8$ eV [panels (a\&b)] and for moderate interactions
$U_d=4$ eV [panels (c\&d)] (including also weaker screening with
$\alpha=0.5$) the correlations induce a gap in the $x^2-y^2$ band
\cite{Lec20}. The lowest states for electrons are then in the $s$ band,
so that going towards the undoped regime involves doping the $s$ band.
Spectra taken at quarter filling lead to analogous interpretations.
This can be seen in Fig.~\ref{fig:mi}, where we show the density of
states for eight electrons (i.e., quarter filling). Data were obtained
by means of TBC, integrating over five sets of boundary conditions.

At strong screening (i.e., for weak $s$-orbital interactions), both
$x^2-y^2$ and $s$ states are occupied and can appear quite close to the
Fermi energy regardless of the value of $U_d$, see
Figs.~\ref{fig:spectra}(a) and \ref{fig:spectra}(c).
Surprisingly, the spectra shown in Figs.~\ref{fig:spectra} depend
little on $U_d$ and stronger on the screening. At strong screening when
$\alpha=0.2$, the occupied states in the $x^2-y^2$ band are rather
similar for $U_d=8$ eV and $U_d=4$ eV, except that the curvature of the
occupied states changes along the $(\pi,0)-(0,\pi)$ line.
Since this implies that interactions $U'$ and $J_H$ between $d$ and $s$
states do here not play a significant role, it supports the notion of a
correlated (and doped) $d$ band that is only affected by a metallic $s$
band via self-doping.

\begin{figure}[t!]
    \vskip .3cm
	\includegraphics[width=\columnwidth]{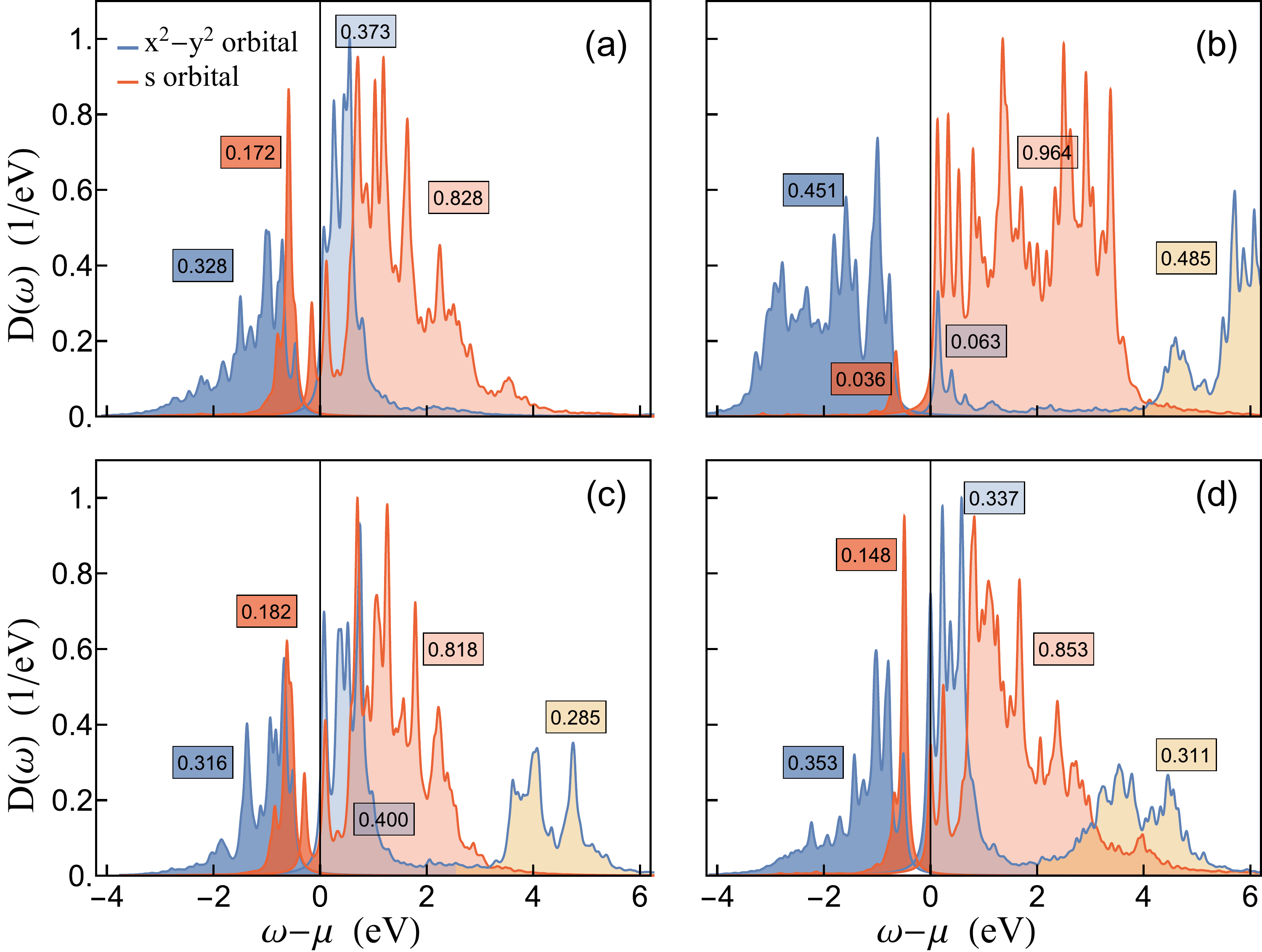}
	\caption{
Density of states $D(\omega)$ of undoped nickelate for weak and moderate
screening with $\alpha=0.2$ and $\alpha=0.5$:
(a)\&(b) $U_d = 8$ eV, and
(c)\&(d) $U_d = 4$ eV.
The conventions are the same as in Fig. \ref{fig:mi}.
}
	\label{fig:dos}
\end{figure}

In contrast, for stronger correlations, i.e., weaker screening
$\alpha=0.5$, spectra shown in Figs.~\ref{fig:spectra}(b) and
\ref{fig:spectra}(d) are affected by $U_d$. All electrons are here in
the correlated $x^2-y^2$ states. It is remarkable that the occupied
states fall almost at the same energies, independently of whether
$U_d=8$ eV or $U_d=4$ eV [cf. Figs. \ref{fig:dos}(a\&b) and
\ref{fig:dos}(c\&d)]. However, splitting between $d$ and $s$ states is
clearly affected by $U_d$ (via $U'$ and $J_H$), which indicates that the
$s$- and $d$-bands are in this regime directly coupled, not only via
self-doping.

Analogous conclusions can be drawn from the undoped density of states
shown in Fig.~\ref{fig:dos}, where Figs.~\ref{fig:dos}(a) and
\ref{fig:dos}(c) are extremely similar: In the regime of strong
screening, both bands are partially filled and results hardly depend on
$U_d$ at all. As already discussed for Figs.~\ref{fig:spectra}(a\&c)
above, this suggests that correlations between the two bands do here
not play a significant role. In the intermediate regime on the other
hand, both bands are likewise partially filled. The comparison with
Figs.~\ref{fig:spectra}(b\&d) indicates that the $s$ states being close
to the Fermi level could be doped away. In this regime, results depend
on $U_d$, indicating that correlations are here more important to
describe low-energy features close to the Fermi level.

Interestingly, the screening increases the density of $s$ electrons and
simultaneously $n_d$ decreases as in the undoped case $n_d+n_s=1$. This
makes the lower Hubbard band (LHB) less than half-filled and
considerable spectral weight is transferred from the upper Hubbard band
(UHB) to the unoccupied part of the LHB (i.e., above the Fermi level
$\mu$ and below the gap). The mechanism of such a spectral weight
transfer is well known in the partly filled Hubbard model
\cite{EskPRL,Esk94} and it explains why the weight of the LHB exceeds
0.500 per spin. Here doping in the Mott insulator is mimicked by the
partial filling of the $s$ band. The largest transfer of spectral weight
is found at $U_d=4$ eV and $\alpha=0.5$, see Table I. The UHB forms only
in the correlated $x^2-y^2$ band and no Hubbard subbands form within
the $s$ band even at $U_d=8$ eV.

Altogether, the densities of states $D(\omega)$ give a metallic regime
for intermediate ($\alpha=0.5$) and strong ($\alpha=0.2$) screening of
strongly correlated $x^2-y^2$ states, see Fig.~\ref{fig:dos}. A large
gap between the Hubbard subbands opens when $U_d=8$ eV; this gap is
reduced to $\sim 0.5$ eV when $U_d=4$ eV. Nevertheless the system has
still a gap which separates separating the Hubbard subbands. The
electronic structure for the $x^2-y^2$ band is typical for a doped Mott
insulator, with the weight of the UHB reduced by the kinetic processes
in a doped system \cite{EskPRL,Esk94}. Indeed, the weight in the LHB
above the Fermi energy increases by $\sim 2\delta$ where $\delta$ stands
for the doping in the LHB, what would also be the weight transferred
from the upper to the LHB by finite doping. In this regime the $s$ band
is only weakly correlated and Hubbard subbands are not visible.

\begin{table}[t!]
\caption{
Electron densities $n_d$ and $n_s$ per spin obtained in the undoped
nickelate for screened interactions $(\alpha<1)$. The weight of the
LHB $w_{\rm LHB}$ in increased by the kinetic weight transfer from
the UHB \cite{EskPRL,Esk94}.
\label{tab:wei}
}
\begin{ruledtabular}
\begin{tabular}{ c c c c c c }
$U_d$ (eV) & $\alpha$ & $n_d$ & $n_s$ & $w_{\rm LHB}^{\;>}$ & $w_{\rm LHB}$ \\
\colrule
8.0 & 0.20 & 0.328 & 0.172 & 0.373 & 0.701   \\
    & 0.50 & 0.451 & 0.036 & 0.063 & 0.514   \\
\colrule
4.0 & 0.20 & 0.316 & 0.182 & 0.400 & 0.716   \\
    & 0.50 & 0.353 & 0.148 & 0.337 & 0.690   \\
\end{tabular}
\end{ruledtabular}
\end{table}


\section{Pairing symmetry}

\subsection{Search for superconducting correlations}
\label{sec:sup}

To investigate pairing symmetries
we calculate ground state overlaps \cite{Mor09,Nic11} between the
undoped ground state $(N=8)$ and the one with two holes $(N=6)$, i.e.,
$\braket{\Phi(N=8)|\Delta_{n}|\Phi(N=6)}$, where $|\Phi(N)\rangle$ is
the ground state for $N$ electrons. Pairing operator $\Delta_{s/d}$
corresponds to $s$- and $d$-wave symmetry,
\begin{eqnarray}
\label{eq:delta_s_d}
\Delta_{s/d}&=&
        \sum_{\substack{i,\mu,\sigma\neq\sigma^{'}\\ \lambda=-1,1}}\!
        d^{\dagger}_{i\mu,\sigma}
        \left(d^{\dagger}_{i+\lambda\hat{x},\mu,\sigma^{'}}\pm
        d^{\dagger}_{i+\lambda\hat{y},\mu,\sigma^{'}}\right),
\end{eqnarray}
where $\mu$ labels $x^2-y^2$ and $s$ orbitals and the $+$ ($-$)
sign refers to $s$-wave ($d$-wave) pairing. $\hat{x}$ ($\hat{y}$)
point to nearest neighbors in $x$ ($y$) direction.
We consider here only intra-orbital
pairs, as we found inter-orbital weight to be negligible.

\begin{figure}[b!]
    \vskip .3cm
	\includegraphics[width=\columnwidth]{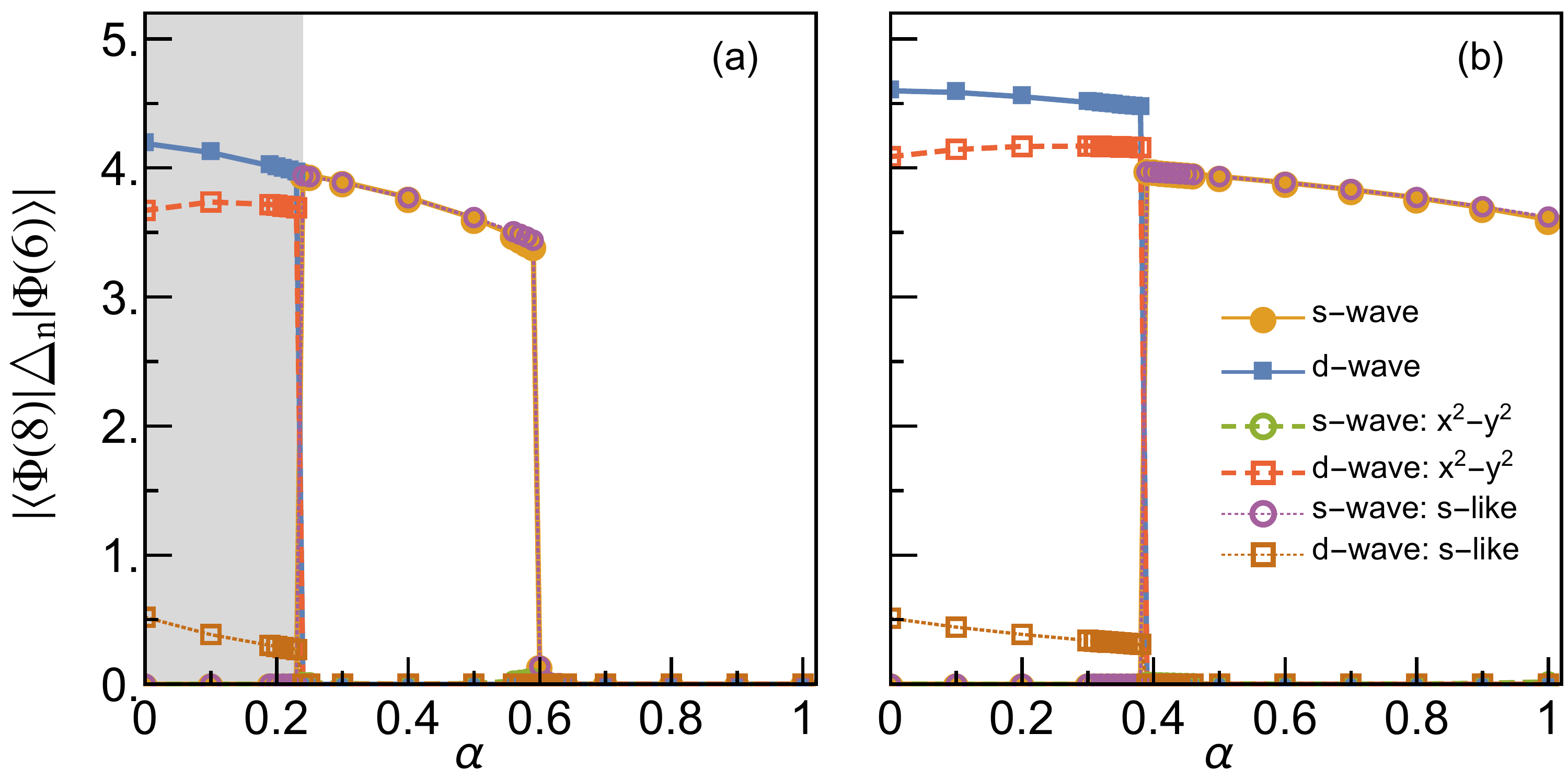}
\caption{Ground state overlap $\langle\Phi(8)|\Delta_n|\Phi(6)\rangle$
          as a function of screening parameter $\alpha$ for:
	  (a) $U_d=8\;\textrm{eV}$, $J_0=1.2\;\textrm{eV}$;
	  (b) $U_d=4\;\textrm{eV}$, $J_0=0.6\;\textrm{eV}$.
          Solid lines indicate  $s$- and $d$-wave
          symmetries; dashed (dotted) line for $x^2-y^2$ ($s$) orbital. The
          shaded region indicates the region with some triplet
          tendencies (see text).}
	\label{fig:overlap}
\end{figure}

For strong interactions $U_d=8\;\textrm{eV}$ and $\alpha >0.6$, where
the undoped ground state is an AF insulator, neither pairing can be
found; the doped holes here prefer to be further apart. In the
intermediate regime, where doped holes were found to prefer the $s$-like
orbital, $s$-wave pairing is found, see Fig.~\ref{fig:overlap}. The
$x^2-y^2$ orbital does here not participate in the pairing. The regime
would thus be best described with a Kondo-lattice like model, where the
$x^2-y^2$ orbital provides spins and the $s$-like band itinerant
carriers~\cite{Hep20,Wan20p,ZYZ20,Bee21}.

\begin{figure}[t!]
	\includegraphics[width=\columnwidth]{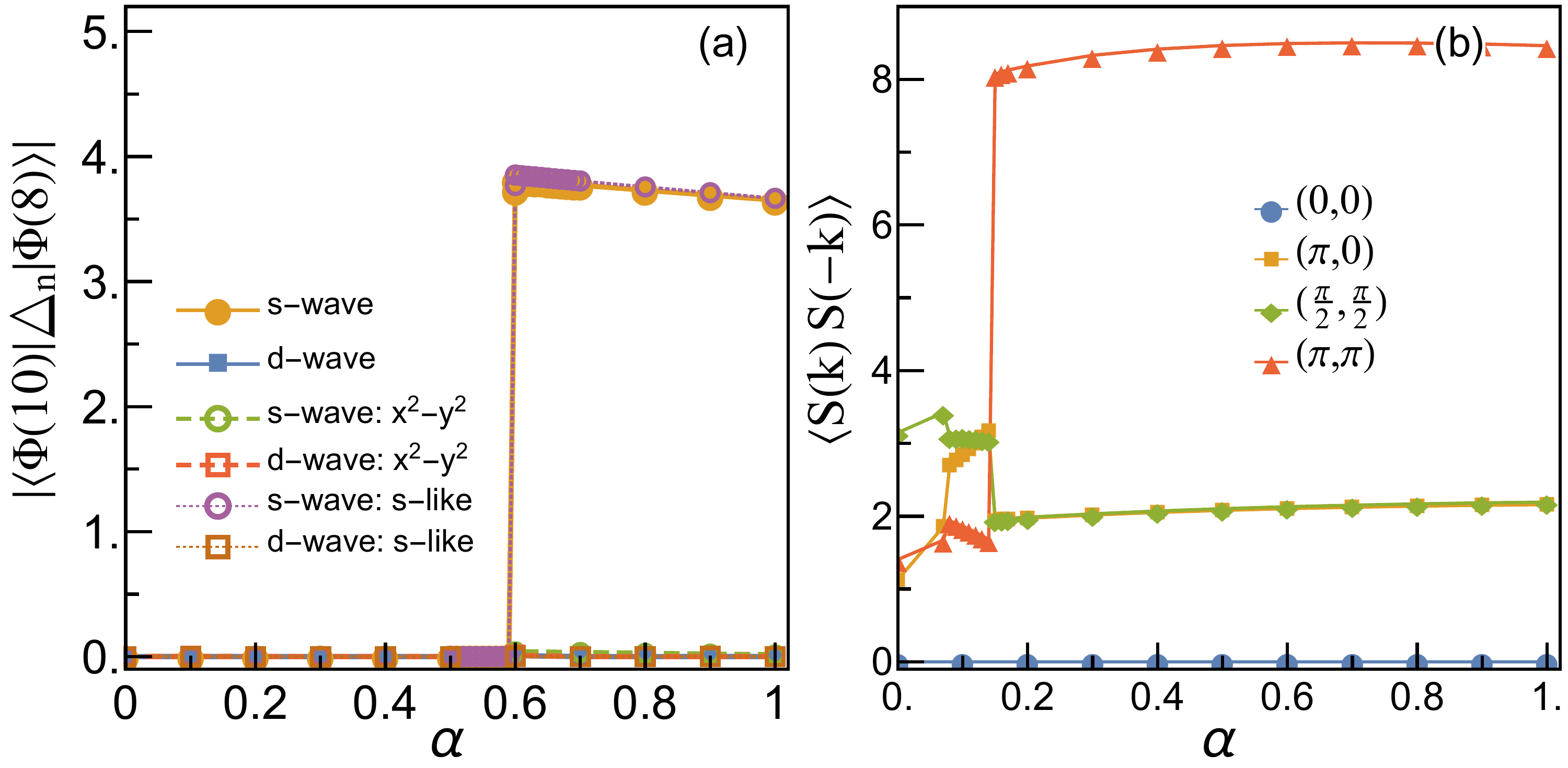}
	\caption{
Ground state of electron-doped superconductors for $U_d=8$ eV, $J_0=1.2$ eV:
(a) Ground state overlap $|\langle\Phi(8)|\Delta_n|\Phi(10)\rangle|$
    as a function of screening parameter $\alpha$ for $s$-wave and $d$-wave
    symmetry, see legend, and
(b) spin structure factor $\langle S(k)S(-k)\rangle$ at selected special
    points of high symmetry, see legend, and for a filling of 10 electrons.
}
	\label{fig:overlap2}
\end{figure}

On the other hand, for strong screening, the pairing is of $d$-wave
symmetry and the involved holes are in the $x^2-y^2$ states. Both the
doped and undoped ground states contain here finite electron density
in the $s$-like band, but their contribution to the pairing is small,
see the open squares in Fig.~\ref{fig:overlap}. Most of the weight is
found in pairs made up of $d_{x^2-y^2}$ holes. In this regime, the
$s$-like band does not play an important role~\cite{Kit20}, its main
effect is to increase hole concentration in the $x^2-y^2$ states via
self-doping.

In addition to hole-doped NdNiO$_2$, electron-doped NdNiO$_2$ is
discussed in Fig. \ref{fig:overlap2}, which shows the overlaps involving
the states with 10 and 8 electrons.	In the strongly correlated regime
$\alpha>0.5$, where the undoped system is an AF with half-filled
$x^2-y^2$ orbital, we find a tendency towards $s$-wave pairs formed by
$s$-band electrons. We find no pairing for stronger screening
$\alpha<0.5$ or for $U_d = 4$ eV, in stark contrast to hole doping. The
reason is that the $x^2-y^2$ orbital of the undoped system is here only
partially filled: extra electrons then enter the $x^2-y^2$ band, see
also the unoccupied states in Figs.~\ref{fig:dos}(a\&b) and
\ref{fig:dos}(c\&d). This moves the $x^2-y^2$ orbital towards half
filling and favors AF order \cite{Ryee20,Jin20,Kita20,Nic20} rather
than superconductivity.

\subsection{Phase diagram}
\label{sec:phd}

\begin{figure}[t!]
	\includegraphics[width=\linewidth]{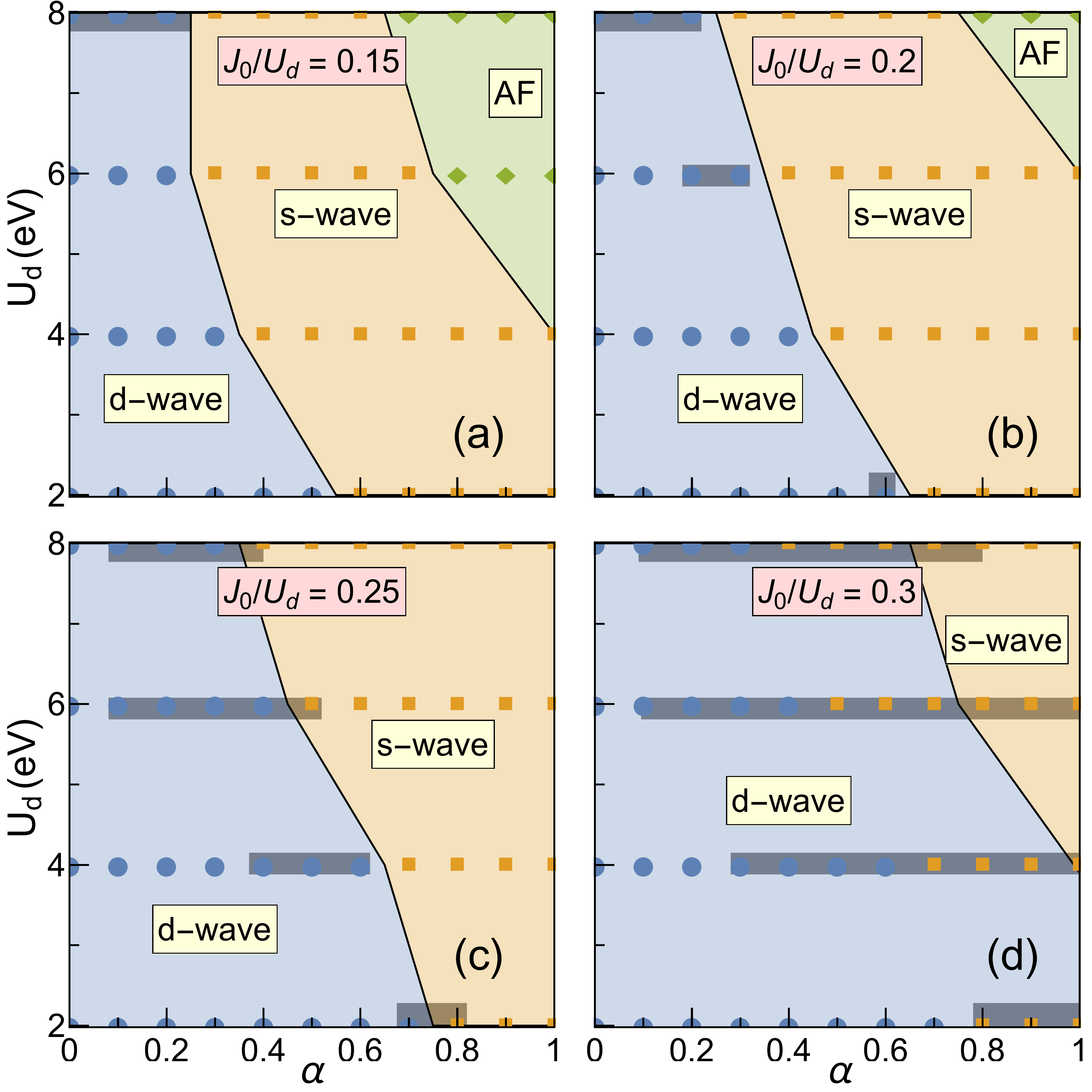}
\caption{NdNiO$_2$ Phase diagram in $(U_d,\alpha)$ plane for increasing
Hund's exchange, i.e., increasing ratio $J_0/U_d$:
(a) $J_0/U_d=0.15$ (as used above); (b) $J_0/U_d=0.2$;
\mbox{(c) $J_0/U_d=0.25$;} (d) $J_0/U_d=0.3$.
Triplet tendencies are indicated by gray shading. For $J_0/U_d\ge 0.25$
in (c)\&(d) AF order vanishes and the phase diagram contains only SC
phases. Note that the lowest value of $U_d$ used (i.e., $y$-axis offset)
is $2\;\textrm{eV}$. }
	\label{fig:phase}
\end{figure}

We now turn to the influence of Hund's exchange coupling and collect
information on pairing symmetries, displayed in the phase diagrams
presented in Fig.~\ref{fig:phase}. For strong and unscreened Coulomb
repulsion and not too strong Hund's exchange coupling, the undoped
system shows AF order and no sign of pairing. In this regime, the
$s$-like band is empty and the $x^2-y^2$ bands is half-filled and Mott
insulating, see Figs.~\ref{fig:density}(a) and \ref{fig:mi}. For weaker
correlations (intermediate screening), AF order is replaced by $s$-wave
pairing (practically only involving $s$-band holes), while $d$-wave
pairing arises at strong screening. In this last regime, some electrons
are found in the $s$-like band, but the doped holes enter the $x^2-y^2$
band, see Fig.~\ref{fig:density}, and
pairing involves mostly the $x^2-y^2$ orbital. Figure~\ref{fig:phase}
illustrates that stronger Hund's-exchange pairing reduces effective
correlations, suppresses AF order, and promotes $d$-wave pairing.

In addition to AF phase and $s$-wave or $d$-wave pairings, we find some
indications of triplet pairing, especially at stronger Hund's exchange
coupling, see Figs.~\ref{fig:phase}(c) and \ref{fig:phase}(d), but also
for very strong bare onsite interaction $U_d=8$ eV, see
Fig.~\ref{fig:phase}(a). Energies obtained for doping with one
$\uparrow$ and one $\downarrow$ hole are here degenerate with the
energies obtained with two $\uparrow$ holes, indicating that the doped
pair is a triplet. In order to check the stability of this result, we
also used TBC. The degeneracy is then lifted and the $S^z=0$ state has
lower energy, suggesting that triplet pairing might be a finite-size
effect. Moreover, the needed Hund's exchange would be rather large
($J_0/U_1\gtrsim 0.25$).

\subsection{LaNiO$_2$ versus NdNiO$_2$}
\label{sec:la}

\begin{figure}[t!]
  \includegraphics[width=\columnwidth]{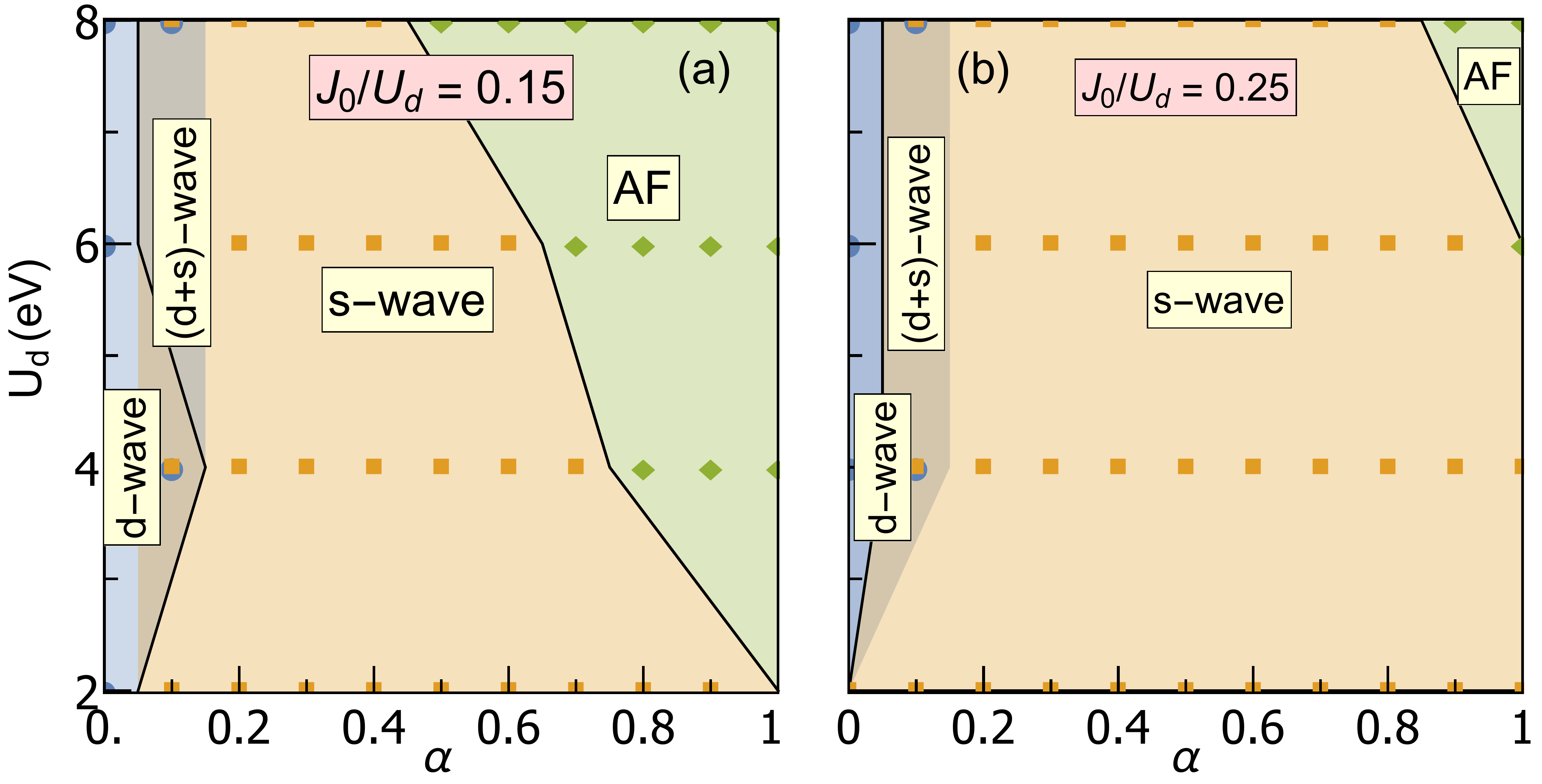}
  \caption{
    Phase diagram for LaNiO$_2$ under hole doping in the
    $(U_d,\alpha)$ plane obtained for: (a) $J_0/U_d=0.15$ and (b)
    $J_0/U_d=0.25$. }
	\label{fig:La_hole}
\end{figure}

\begin{figure}[b!]
  \includegraphics[width=\columnwidth]{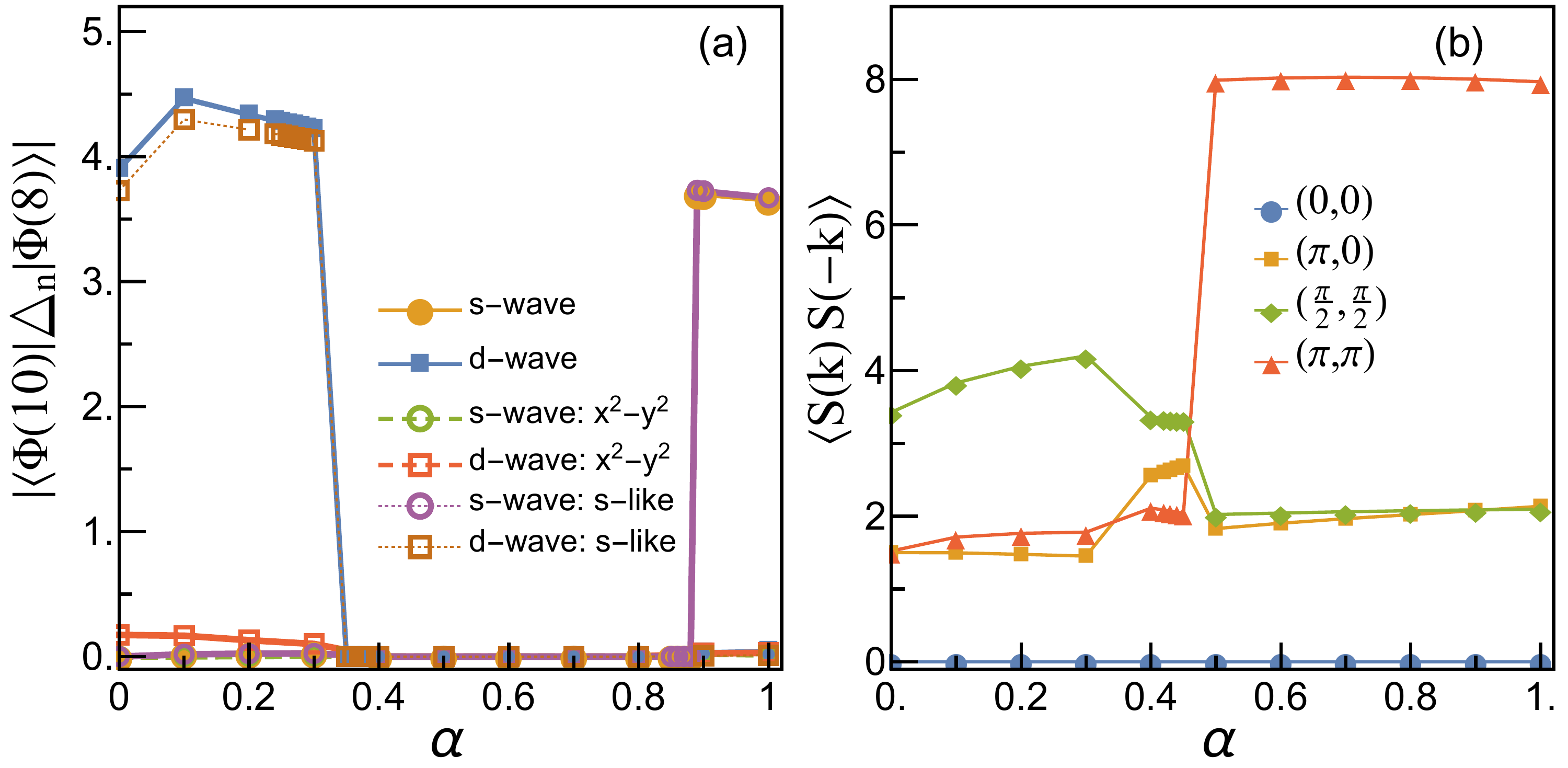}
  \caption{
    Electron doping for LaNiO$_2$ depending on screening $\alpha$ and
    for $J_0/U_d=0.15$, $U_d=4$ eV. The panels show:
    (a) ground state overlap with two electrons added to quarter
    filling and (b) Spin structure factor for 10 electrons.}
  \label{fig:La_el}
\end{figure}

Furthermore, the electronic properties of LaNiO$_2$ compound are obtained
starting from the electronic structure
\cite{Jia19,Si20,Wu20,Kle21,Zha21,Bee21,Hig21,Bot22,Zha21a,Isl21,Zen21}.
The model parameters are provided by Ref. \cite{Adh20}. Surprisingly,
the hole-doped phase diagram shows largely $s$-wave pairing as well as
an AF phase (see Fig. \ref{fig:La_hole}). Some $d$-wave pairing is also
found, however, it requires strong screening strength
$\alpha\approx 0.1$ to develop. AF order is robust and remains stable
even at $U_d=4$ eV, highlighting the importance of rare-earth screening
required for superconductivity in LaNiO$_2$. The smaller $\Gamma$-pocket
\cite{Adh20} implies pairing toward $s$-wave, but its robustness against
$U_d$ is not expected and this type of order is more subtle. Compared to
NdNiO$_2$, the parameter range for $d$-wave pairing is strongly reduced
in LaNiO$_2$, and $s$-wave pairing dominates the phase diagram. In
addition to $d$- and $s$-wave SC phases, we observe non-vanishing of
$(d+s)$-wave at the crossover between the two above symmetries. Although
this exotic type of pairing is interesting, with the small cluster size
used here we cannot argue that it is stable in the thermodynamic limit.

Figure~\ref{fig:La_el} shows pairing overlaps and spin-structure factor
for electron-doped LaNiO$_2$. As for electron-doped NdNiO$_2$
(see Fig.~\ref{fig:overlap2}), strong correlations enhance AF order in
the $x^2-y^2$ orbital and allow coexisting $s$-wave pairing in the
$s$-like band. At weak correlations, however, we now find $d$-wave
superconductivity in the regime without AF order, in contrast to
electron-doped NdNiO$_2$. However, in stark contrast to the hole-doped
scenarios, the $d$-wave pairs are composed almost exclusively of $s$
electrons rather than $x^2-y^2$. This result suggests that
superconductivity in electron-doped LaNiO$_2$ would be possible.

\section{Summary and conclusions}
\label{sec:summa}

\textcolor{black}{We have used exact diagonalization to investigate an
 effective two-band model for infinite-layer nickelates, where the band
 with a strong Ni($d_{x^2-y^2}$) character can be expected to be more
 correlated than the one with a rather extended $s$-like wave function
 of mostly rare-earth character.}
We focus here on the interactions in both bands, especially their
relative strength, which also tunes the highly relevant~\cite{Adh20}
interorbital interactions between the two orbitals.
The latter give interband interactions and are
responsible for the pairing.

On both ends, the very strongly correlated and the strongly screened
regimes, the two-band model can be mapped onto a single Hubbard-like
band. For (unrealistically) strong interactions, we find an AF Mott
insulator without tendencies to superconductivity. In the more
realistic screened regime, the $s$-like band takes up some of the
charge carriers, which can easily be accounted for effectively by
adjusting the doping level of the correlated $x^2-y^2$ band~\cite{Kit20}.

For intermediate screening, in contrast, the $s$-band hosts the doped
holes forming $s$-wave pairs, so that it cannot be neglected.
The situation broadly corresponds to a Kondo-lattice--like scenario,
with the caveat that the 'localized' $d_{x^2-y^2}$ spins can also move
\cite{Hep20,Wan20p,ZYZ20}. Hund's exchange coupling naturally yields
ferromagnetic interaction between itinerant $s$ carriers and
$d_{x^2-y^2}$ spins, but it is interesting to note that $s$-wave
pairing at stronger coupling was also obtained in a similar effective
model with AF spin-spin coupling~\cite{Wan20p}.

\textcolor{black}{Hopping parameters used were modeled on NdNiO$_2$,
  and we find that a self-doped $x^2-y^2$ band can likely capture some
  regimes of a model for hole-doped NdNiO$_2$, while the $s$-like band
  would be expected to play a stronger role at electron doping.
  We also used a slight modification of the model in order to arrive
  at a model more appropriate to LaNiO$_2$ and conclude that the
  $s$-like band can be similarly be expected to play a stronger role.}
Electron doping enhances antiferromagnetism, and  $s$-wave pairing due
to the $s$ orbital might then arise, while $d$-wave pairing is only
found for LaNiO$_2$, but not for NdNiO$_2$. We find that the overall
phase diagrams are similar for
\textcolor{black}{model parameters aiming at} NdNiO$_2$ and LaNiO$_2$.
However, a mapping onto a single Hubbard band is here applicable to a
significantly reduced part of the parameter space in the case of
LaNiO$_2$, with a much broader regime falling into the Kondo-lattice---
like two-band regime. We thus conclude that many, but not necessarily
all, aspects of Ni-based superconductors can be discussed in an
effective one-band scenario, in agreement with Ref.~\cite{Lec21}.

Experimental evidence is on pairing symmetry is at the moment not
completely clear. Recent observation on both Nd- and La-based compounds
suggests isotropic nodeless pairing in Nd-nickelate while it is
anisotropic nodeless or nodal + nodeless pairing in La-nickelate
\cite{Chow22}. Tunneling spectra in Nd$_{1-x}$Sr$_x$NiO$_2$ thin films
have revealed $d$- as well as $s$-wave gaps~\cite{Gu20a}. Different
surface termination has been conjectured to underlie this observation
and our results might provide a rationalization: if termination reduces
screening locally, it can push the system into the $s$-wave regime.

\textit{Note added.}---After this work was completed, we became aware
of a recent two-band model study of nickelates by the Stanford group
\cite{Pen21}. Common feature is the coexistence of a strongly
correlated $x^2-y^2$ orbital and a weakly correlated $s$-like orbital
which supports the relevance of such a two-band model for the
superconducting infinite-layer nickelates.

\acknowledgments

We thank Wojtek Brzezicki, Andres Greco, and George A. Sawatzky
for very insightful discussions. T.~Plienbumrung acknowledges
\mbox{Development} and Promotion of Science
and Technology Talents Project (DPST). A.~M.~Ole\'s acknowledges
Narodowe Centrum Nauki (NCN, Poland) Project No. 2016/23/B/ST3/00839
and is grateful for support via the Alexander von Humboldt Foundation
\mbox{Fellowskip} \mbox{(Humboldt-Forschungspreis).}

\bibliography{twobas}

\end{document}